\documentclass[runningheads,a4paper]{llncs}

\usepackage{amssymb}
\usepackage{graphicx}
\usepackage{url}
\newcommand{\keywords}[1]{\par\addvspace\baselineskip
\noindent\keywordname\enspace\ignorespaces#1}

\begin{document}

\mainmatter  

\title{The Analogue Computer as a Voltage-Controlled
Synthesiser}

\titlerunning{Analogue Computer as a VCS}

%
%
\author{Victor Lazzarini and Joseph Timoney}
%
\authorrunning{V. Lazzarini and J. Timoney}

\institute{Maynooth University, \\Maynooth, Co. Kildare, \\Ireland \\\email{\{victor.lazzarini,joseph.timoney\}@mu.ie}}

%
%

\maketitle

\begin{abstract}
This paper re-appraises the role of analogue computers within electronic and 
computer music and provides some pointers to future areas of research. It begins by introducing the
idea of analogue computing and placing in the context of sound and music applications.
This is followed by a brief examination of the classic constituents of an analogue
computer, contrasting these with the typical modular voltage-controlled synthesiser.
Two examples are presented, leading to a discussion on some parallels between these two
technologies. This is followed by an examination of the current state-of-the-art in 
analogue computation and its prospects for applications in computer and electronic music.
\end{abstract}

\section{Introduction}\label{introduction}

Computer Music, for the most of its history, has been concerned with the
use of what we can generally class as the \emph{digital stored-program
computer}, although this (correct) terminology has by now fallen in
disuse, due to the ubiquitous nature of these devices. In this paper, we
will instead look at a different type of computer and its potential to
sound and music design, and its relationship to the music instrument
technology of voltage control. The principles we will explore fall into
the category of \emph{analogue computing}, which approaches both the
actions involved in computation, the modelling and the problem design
from a different perspective.

The principles that constitute analogue computing can be seen from two
perspectives that are somewhat independent from each other. On one hand,
the hardware that implements it allows for the solution of problems
containing continuous variables, whereas digital circuitry implies a
discretisation of these (even in models that assume underlying
continuous quantities). In the case of music, the possibility of time
and amplitude-continuous computation is significant, considering the
amount of work that has been dedicated to solving discretisation issues
in areas such as virtual analogue models \cite{Pakarinen}.

From a different perspective, analogue computing approaches problems in a
way that largely dispenses the algorithmic approach of digital computer
programming in favour of hardware reconfiguration, setting up the
computation not so much as a sequence of steps, but as an
interconnection of components \cite{Ullmann}. This also implies that the
hardware model set up in this way is an \emph{analogue} of the problem
at hand. Additionally, while analogue computers may be able to compute
problems related to the steady state of a system, they are more
frequently used to providing solutions relating to transient behaviour
\cite{Navarro}. Such problems are significant to sound and music applications,
where the dynamic properties of a system are fundamental.

Analogue computing has had a long history, which began with mechanical
devices that were used as aids to calculation of specific problems
(navigation, gunnery, accounting, etc.), and became a major scientific
field of research with the advent of practical electronic devices. These
could be combined more flexibly to realise various types of modelling.
From a music perspective, these developments influenced the technology
of voltage control, and the modular aspect of electronic analogue
computers appears to be significant in providing the priciples
underpinning early synthesisers \cite{Moog}.

In this paper, we will examine the relationships that exist between
these devices. We will start by exploring the principles of analogue
computation with electronic computers. This will be followed by an
introduction to modular voltage-controlled synthesisers from the
perspective of analogue computing. Then we will examine the
possibilities of general-purpose electronic computers as musical
instruments, followed by an examination of the current state of the art
in the area and the perspectives for new research in sound and music
(analogue) computing.

\section{Electronic Computers}\label{electronic-computers}

Analogue computers, as discussed in the introduction to this paper,
operate under different principles to their digital stored-program
counterparts. Generally, they are set up to provide solutions to a
problem that is laid out in terms of a mathematical equation or set of
equations, providing an answer to these, given a certain input. In this
case, the type of problems that are applied to them can be of different
characteristics, provided that they can be described in an algebraic
form. Programming the computer is then a matter of setting an analogue
to the original problem \cite{Ullmann} by means of various computing elements.
Therefore, the capabilities of a given analogue computer are determined
by the types of computing blocks it can offer, and how they can be
connected in a program.

\subsection{Computing elements}\label{computing-elements}

Analogue computers are made of various types of components that often
operate as \emph{black boxes}, providing an output given a set of inputs
and conditions, within a certain level of tolerance. The inputs and
outpus of such boxes are electric signals whose voltages play the part
of the variables that are manipulated in a program. Programs will then
be made up of patching connections between these different blocks,
setting up the initial conditions that configure the problem and then
running the computer, from which the answer or answers can be read by
appropriate output devices.

While the components of an analogue computer can be quite varied in
nature, there are some key blocks that are present universally in these
devices, to provide basic computing operations.

\paragraph{Arithmetics}\label{arithmetics}

\begin{figure}[htp]
\centering
\includegraphics[width=0.25\columnwidth]{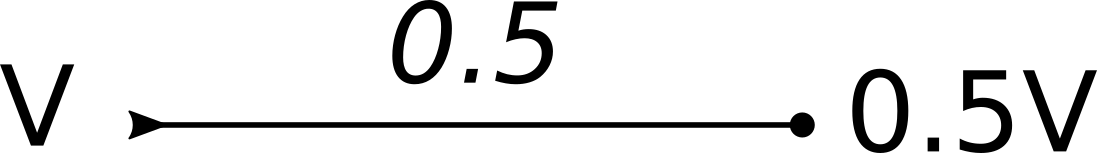}
\caption{Attenuation example}
\label{fig:atten}
\end{figure}

We can divide the arithmetic operations into three fundamental
categories, that are addressed by specific types of electronic circuits:
(a) multiplication by a scalar; (b) addition/sum; (c) multiplication of
signals. In the case of (a) and (b), a fundamental component is the
\emph{operational amplifier} \cite{Ragazzini}. This component allows a gain to be
applied to the signal, and facilitates both multiplication and addition
to be implemented. Of course, if only attenuation is required, then a
signal can be passively modified by a variable resistance (fig. \ref{fig:atten}), but
in all other cases, the op amp is required.

\begin{figure}[htp]
\centering
\includegraphics[width=0.5\columnwidth]{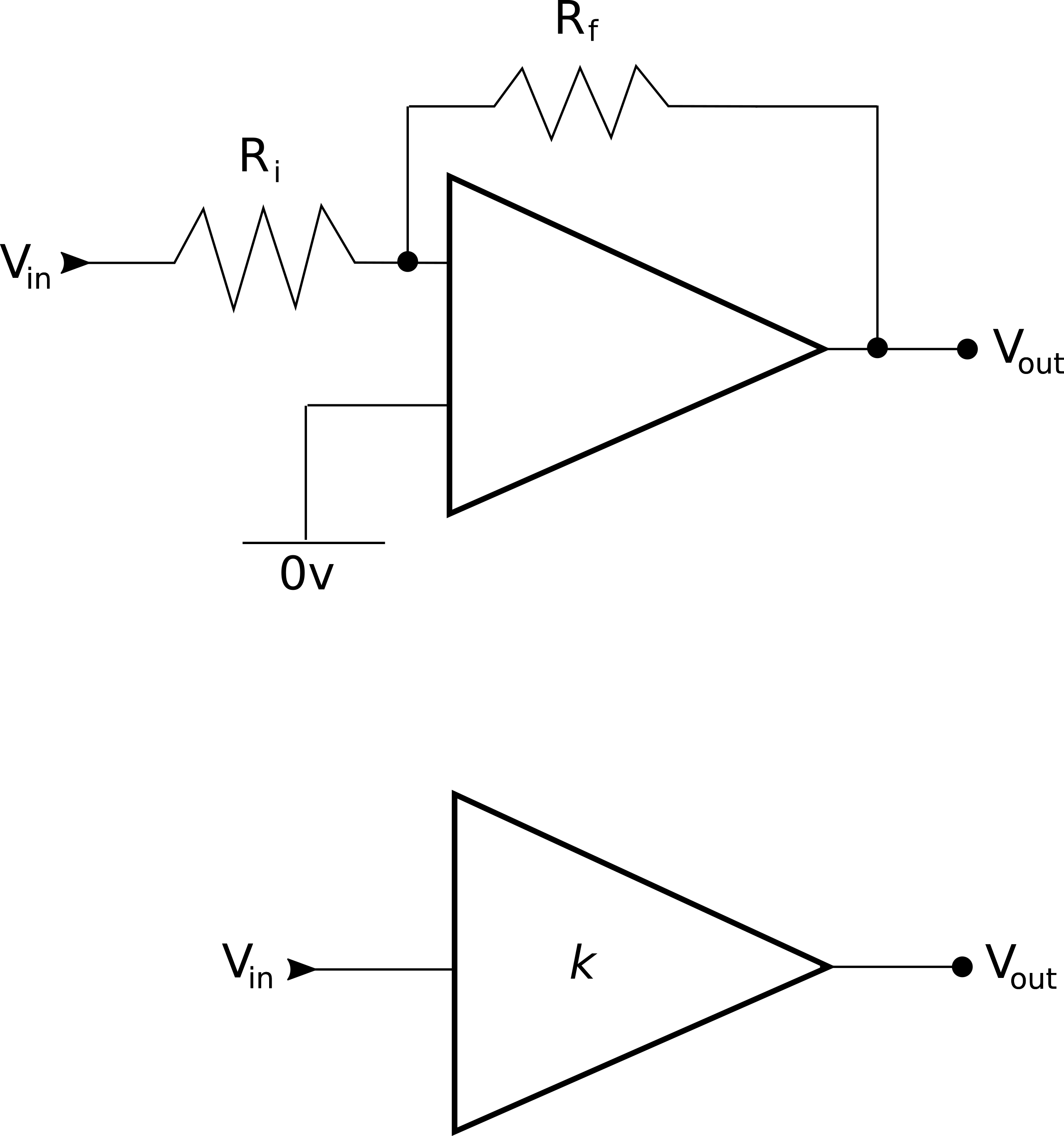}
\caption{Op amp circuit schematics and gain scaling symbol}
\label{fig:opamp}
\end{figure}

Gain scaling is implemented simply by setting the multiplier constant
\emph{k} in the op amp, which is the ratio of the resistances
$R_f/R_i$ that are employed in the circuit (fig. \ref{fig:opamp})

\begin{equation}
V_{out}(t) = -k V_{in}(t)
\end{equation}

Note that the op amp will normally have the effect of inverting the sign
of the voltages applied to its input, due to the fact that only its
inverting input is used.

Summing two voltages also require an op amp (fig. \ref{fig:sum}), and the input
signals are scaled by the ratios of the individual input resistances and
the feedback path resistance, $k =  R_f/R_n$. Note that adding
units such as these can be set up for more than two inputs.

{\begin{figure}[htp]
\centering
\includegraphics[width=0.5\columnwidth]{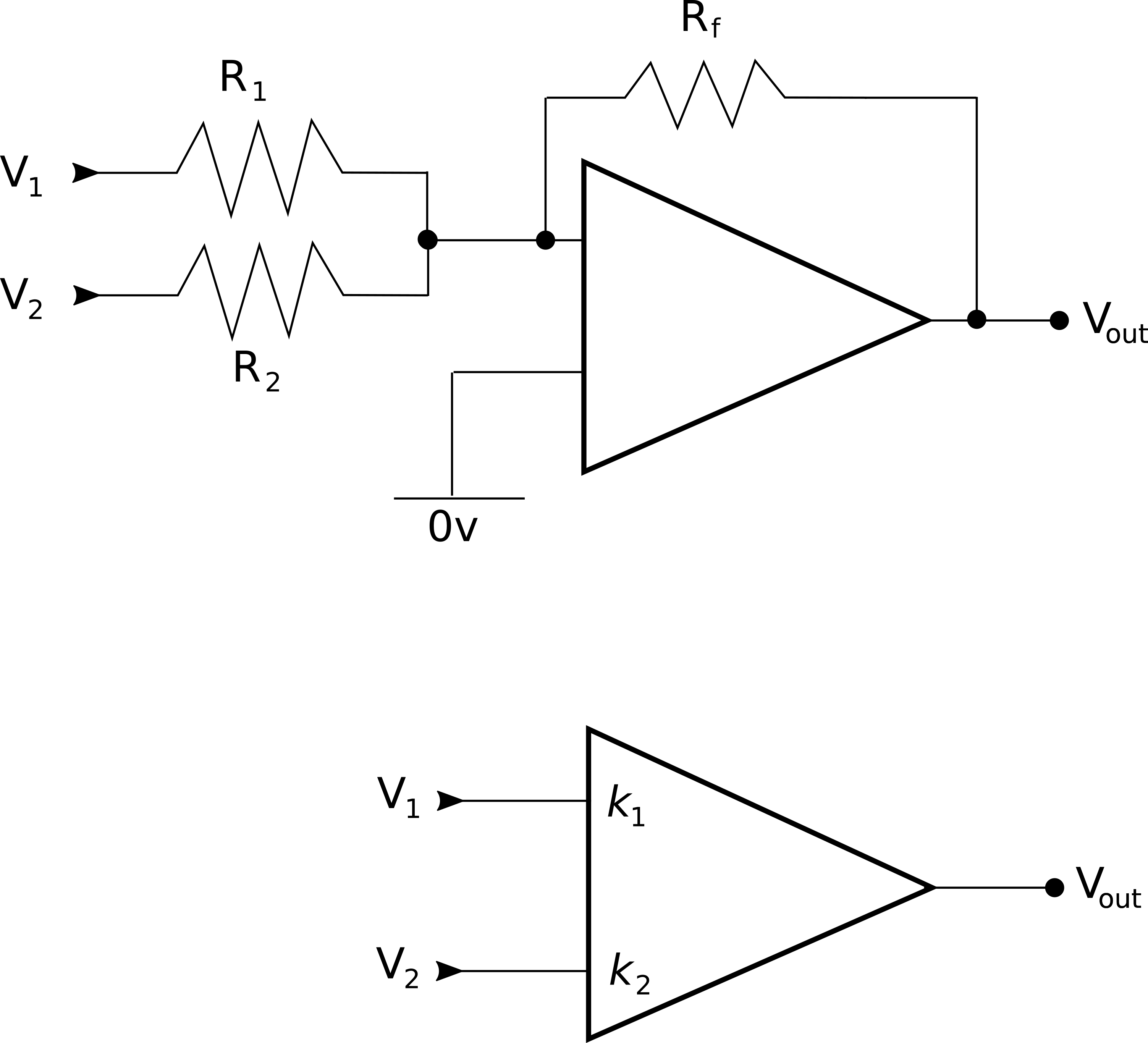}
\caption{Summing amp circuit with two inputs and symbol}
\label{fig:sum}
\end{figure}

\begin{equation}
V_{out}(t) =  - \sum_0^n k_n V_n(t)
\end{equation}

Multiplication of two signals is generally taken as a separate category
as it requires more complex circuitry. In this case, the output is
equivalent to the instantaneous value of the multiplication of its
inputs, scaled by a constant.

\paragraph{Integration}\label{integration}

Another key component of an analogue computer is the integrator. The
means of integrating an input signal is provided by a capacitor, and the
circuit (fig. \ref{fig:integ}) also includes an op amp to complement it. As we can
see, the capacitor replaces the feedback resistor in a simple scalar
multiplier. The output is also scaled by $k = 1/R_iC$ where C
is the capacitance in the op amp feedback path. The voltage across the
capacitor can also be set as an initial condition $V_0$.

\begin{figure}[htp]
\centering
\includegraphics[width=0.5\columnwidth]{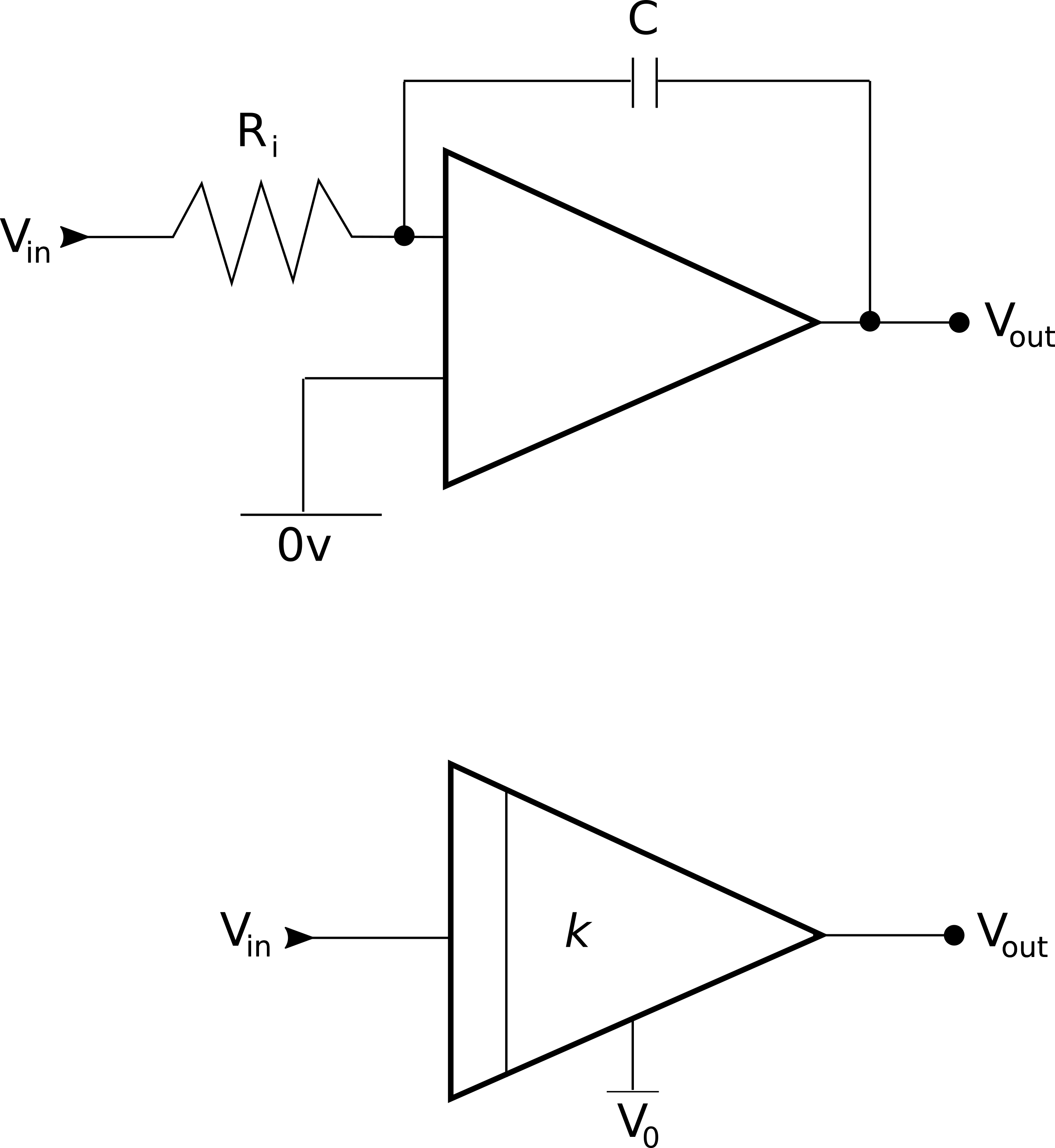}
\caption{Integrator circuit and symbol}
\label{fig:integ}
\end{figure}

\begin{equation}
V_{out}(t) = -k \int_0^t V_{in}(t) + V_0
\end{equation}

It is also a simple matter to include multiple input signals to an
integrator, using a combination of the circuits of figs. 2 and 3. In
this case, the different inputs are scaled and added together before the
integration is performed.

\paragraph{Functions}\label{functions}

It is also fundamental for analogue computers to be able to provide
means of generating a variety of functions. Among these we will find the
usual single-variable functions trigonometric, exponential, triangle,
rectangular, ramp, etc. Some computers would also have more
sophisticated means of generating user-defined functions \cite{Ullmann}. It is
worth noting that function generators is a general class of modules that
also include the multiplication, summation, and integration blocks
described above \cite{Navarro}.

\paragraph{Other Modules}\label{other-modules}

Various other modules exist in various analogue computing devices
\cite{Ullmann}. Logic blocks such as comparators allow voltages to be compared
for binary decisions (such as opening and closing signal connections)
and step-function implementations. Limiters are special-purpose
comparators that can keep signals within a given range. Time delays
provide a means of shifting the phase of functions and can be
implemented in discrete capacitor circuits called bucket brigade devices
\cite{Sangster} Output of analogue computations involve some means of measuring
the voltage of a program, which can be done by various means such as
strip-chart and \emph{xy} recorders, oscilloscopes, voltmeters, and
similar components. A sound and music computing relevant output
block would consist of an audio pre-amplifier that allows line-level connections
to a mixer and power amplifier. This is of course, the main output component
of a voltage-controlled synthesiser.

\section{Modular Synthesisers}\label{modular-synthesisers}

Modular electronic sound synthesis, especially the variety involving
voltage control technologies, has been a cornerstone of electronic music
since the post-war era \cite{Wells}. In examining the devices that have been
and are currently used for this purpose, we can see very clear parallels
with the technology of analogue computers. In fact, analogue
synthesisers in general have been identified as special purpose
computers \cite{Ullmann}. As in that type of technology, modules play an
important part as the components of programs, which in the case of the
typical modular design are made up of patch-cord connections between
them.

\subsection{Modules}\label{modules}

Voltage-controlled synthesizer modules are generally built at a higher level of operation
if compared to analogue computing blocks. This means that access to
fundamental aspects of computation are less common. For example,
function generators in the form of oscillators realise compound
operations, which in the case of an analogue computer would be provided
in smaller building blocks of a time function generator, plus
multipliers and adders. However, some basic elements are given: ring
modulators implement two-input multiplication, mixers are summing
modules, and offset/scaling modules can provide addition and
multiplication of signals and scalars. The typical modules of a synthesiser
include voltage controlled filters (VCFs), oscillators (VCOs),
and amplifiers (VCAs).

In general, modular synthesisers provide a rich set of components, many
of which can be seen as different types of function generators: for
attack-decay-sustain-release curves, noise sources, and sequencers,
which provide user-defined functions. However, the synthesiser set of
components is provided with significant specialisation, and in general
lacks access to the fundamental building blocks of computation. Tracing
an analogy to the music languages used for programming digital
computers, modular synthesizers provide high-level unit generators
\cite{Lazzarini:13}, but not the means of coding (or in this case, setting up) the
unit generators themselves (the modules in the analogue domain).

\section{Examples and Discussion}\label{examples-and-discussion}

In order to the expand the discussion of analogue computing for sound
and music, it is interesting to consider some examples to illustrate
simple operations. While we would put these problems from a
general-purpose computing perspective, we would also like to consider
them with respect to typical sound synthesis applications.

\subsection{Linear Functions}\label{linear-equations}

The simplest example of the application of analogue computation is to
set up the solution to a linear problem, such as

\begin{equation}
f(t) = ax(t) + b
\end{equation}

\noindent which may be applied, for instance, to glide the pitch
of a tone from one frequency to another. The program for this is
shown in figure \ref{fig:linear}, smoothly sliding by a user-defined interval.
In this case, each increment of 1V in the input starting from a voltage  $V_0$, 
will provide a jump of $k$ semitones, when used as a 1V/oct exponential 
frequency signal. 

\begin{figure}[htp]
\centering
\includegraphics[width=0.7\columnwidth]{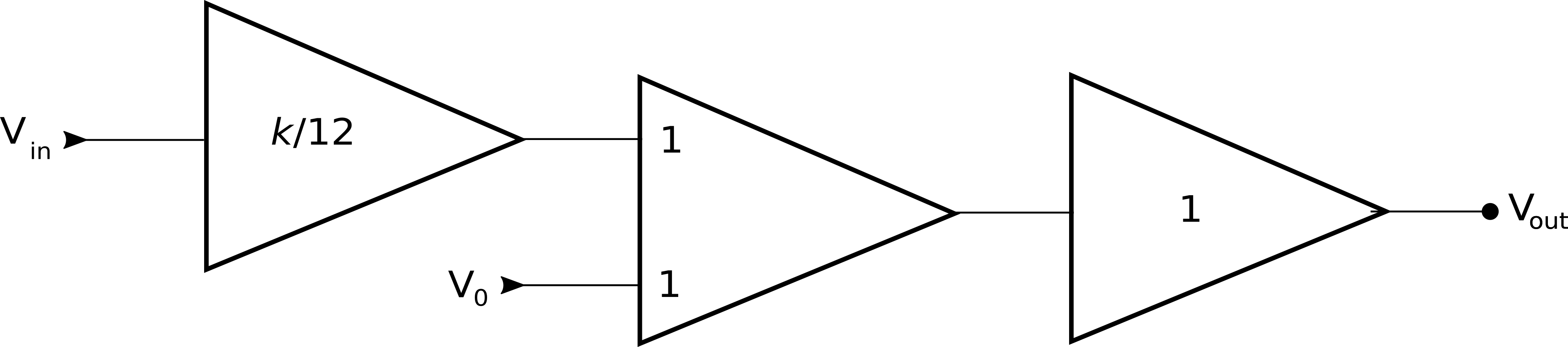}
\caption{Linear equation program}
\label{fig:linear}
\end{figure}

This, of course, can easily be set up in a modular synthesiser by
the use of an amplifier and an offset, which are blocks that are
readily available. At this level of simplicity, the synthesiser can
match the analogue computer almost on a one-to-one component
basis.

\subsection{Differential Equations}\label{differential-equations}

A more common application of analogue computers has to do with the
solution of differential equations. Consider the following example,

\begin{equation}\label{eq:diff}
y(t) = ax(t) - b \frac {d y(t)} {dt}
\end{equation}

\noindent which is a simple first-order differential equation. This can
be translated into an analogue computer program as shown in
fig. \ref{fig:diff}. The significance of this is that such a differential
equation also implements a simple infinite impulse response 
low-pass filter. With this approach, could implement filters of different 
designs, more complex of course, but using the common
blocks of an analogue computer. 

\begin{figure}[htp]
\centering
\includegraphics[width=0.7\columnwidth]{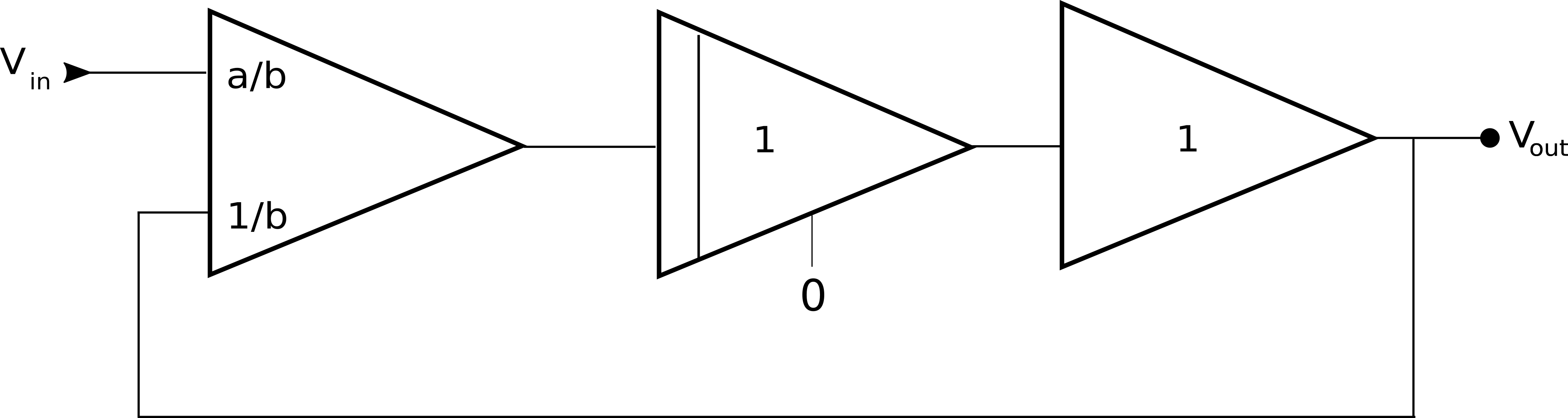}
\caption{Difference equation program}
\label{fig:diff}
\end{figure}

Hutchins \cite{Hutchins} pointed out that the fundamental voltage-controlled 
building block used here (fig. \ref{fig:diff}), the integrator, is applicable to 
the well-known state variable filter (SVF) design.
He noted that the fact that the SVF is formed from elemental blocks of 
two integrators and a summer in a  loop has been known from 
analogue computer programs, where it was used in physical simulations of second-order 
responses by producing voltages corresponding to the magnitudes of the
\emph{state variables} of the system.

This first-principles approach is contrasted now with the filter modules implementing
different topologies, with various particular characteristics,
that are found in voltage controlled synthesisers. The significant difference is that
these are fixed to a given design, and do not allow access or
manipulation of its circuit connections. This demonstrates an
example where there is no one-to-one match between an 
analogue computer and a synthesiser.

\subsection{The General-purpose Analogue Computer as a Musical
Instrument}\label{the-general-purpose-analogue-computer-as-a-musical-instrument}

Given the examples discussed above, it might be surprising to see that
not a lot has been made in terms of utilising general-purpose analogue
computers in musical applications. The composer Hans Kulk appears to be
a solitary figure working in this field \cite{Kulk}. This may be attributed
to various factors: cost, as in the heyday of analogue computers, it was
very expensive to access these; complexity, programming analogue
computers required significant technical expertise, which was not
mitigated by music-directed approaches, as for instance provided by
music programming languages in the case of digital computers.

The existence of the modular synthesiser, in fact, can be seen as the
analogue counterpart to the digital computer music programming
environments. However, as noted above, they were not developed to
provide lower-level access to computation, which is the case of digital
computer programming (as in, for instance, Csound \cite{Csound2016}. Finally,
also we should consider that the obsolescence of the general-purpose
analogue computer, in parallel with the ubiquitousness of the digital
computer, also played a part in the process. However, some new prospects
in the analogue computing domain may allow us to re-appraise these
devices as possible vehicles for music making. The main consideration is
that there is no fundamental impediment to this; on the contrary, there
seems to be fertile ground for work in this respect.

\section{Prospects for Electronic and Computer Music}\label{future-prospects}

While analogue computers may appear to some to have passed their heyday and be 
generally obsolete today, that does not seem to be the case if we look at
some cutting-edge research in the field. The direction of travel seems to 
involve the development of very large scale integration (VLSI) components that
implement a collection of computing elements. These can then be made available
for programming in flexible ways and used as dedicated processors within
a larger host system \cite{Cowan}. The technology of Field Programmable
Analogue Arrays \cite{Lee1,Lee2,Lee3} has also opened some interesting possibilities in the particular
case of analogue signal processing. 

It is within this context that a number of prospects for musical signal processing may arise. The
interest in re-creating analogue environments using digital simulations
that sparked virtual analogue model research may have exhausted its
potential for further refinements, and work that is actually directed to
the \emph{real} thing is an attractive proposition. Music production is one of the few areas of 
technology where analogue signal generation and processing techniques 
continue to be used, and in fact, have enjoyed a significant resurgence

\subsection{Field Programmable Analogue Arrays}
 
Recent developments in analogue signal processing technology included 
the development of Application-specific Integrated Circuits (ASICs),
used in the implementation of synthesiser modules such as filters and oscillators. 
However,  as the name implies,  ASICs target specific purposes and re-designing them is very expensive. 
What would be more suitable is  to have an analogue equivalent of the digital Field Programmable Gate Array (FPGA). 
Fortunately, the concept of Field Programmable analogue Arrays (FPAAs) was introduced with the promise 
that it facilitated analogue components to be connected together in an arbitrary fashion, 
allowing for rapid testing and measurement of many different circuit designs. 

A similar but less sophisticated technology is the PSoC (programmable system-on-chip) by Cypress Semiconductor. 
These chips include a CPU core and mixed-signal arrays of configurable integrated analogue and digital peripherals. 
The FPAA was introduced in 1991 by Lee and Gulak \cite{Lee1}. The idea was further enhanced by the same authors in 1992 \cite{Lee2} and 
1995 \cite{Lee3} where op-amps, capacitors, and resistors could be connected to form a biquad filter, for example. 
In 1995, a similar idea, the electronically-programmable analogue circuit (EPAC) was presented in \cite{Pierzchala}. 

Within the FPAA explored in \cite{nease2014a} it was organised into three functional blocks: 
(1) the computational analogue block (CAB), which is a physical grouping of analogue circuits that act as computational elements, 
(2) the switch matrix (SM), which defines the interconnection of CAB components, and 
(3)  the programmer which allows each device to be turned completely on, turned completely off, or operated somewhere in-between. 
This flexibility means that switch elements can be used for computation as well as routing. 
This is especially beneficial in audio applications, since transistors set to a constant bias are often necessary. 

Only in recent years have FPAAs become powerful enough to be considered for facilitating complex analogue sound
synthesis, in the implementation of common modules. Two papers investigated whether FPAAs were capable of creating entire synthesis systems. 
One paper illustrated how the low-pass VCF developed and popularised by Robert Moog \cite{Moog2} could be implemented using an FPAA  \cite{nease2014a}. 
For this implementation it was found that the FPAA could support 12 VCFs, assuming perfect utilisation of all the available 
resources by the CAB, but it may also be possible to include another 8-10 filters under alternative constraints. A second paper looked 
at the FPAA configuration for a VCO and VCA, and two common control modules, the low-frequency oscillators and 
the envelope generator, which would allow for the development of a complete synthesiser \cite{nease2018applications}. 
The paper identified a number of challenges, which included whether the VCO implementation would be controllable 
over a wide pitch range and remain stable with temperature changes. In general, FPAAs appear to be one of the most
promising analogue technologies for sound and music computing.

\subsection{Programmability}\label{programmability}

Programming a modular synthesiser with patch-cords
 is a significant undertaking, and the traditional analogue
 computers presented much bigger challenges in that
 respect. Clearly, if we are to be able to implement signal-processing
 operations from first principles, the question of programmability takes
 is a key concern. In modern VLSI-based systems such as the one
 described in \cite{Cowan} and in the case of FPAAs, some means of setting up connections 
 between components is provided  (and storing/retrieving these), 
 which can be at various levels. 
 
 We could trace a parallel with digital computers, where the code may be represented by an
 assembly-type language for a given hardware, or in a high-level
 language such as C. In an analogue computer, we could manually 
 translate equations (such as for instance eq. \ref{eq:diff}) into the actual
 hardware connections (e.g. fig. \ref{fig:diff}), or potentially use
 an algebraic compiler to synthesise the necessary circuits (such as 
 the one discussed in \cite{Achour}). We can hypothesise that such
 a high-level language could be developed targeting the requirements
 of analogue signal processing for music computing applications.
 
\subsection{Hybrid Digital-Analogue Systems}\label{hybrid-digital-analogue-systems}

Another emerging characteristic of modern analogue computing appears
to be the development of hybrid digital-analogue systems. One such 
arrangement is described in \cite{Guo}, where a combination of
digital and analogue circuits are used to construct a programmable
device. This type of arrangement is mirrored in modern polyphonic
analogue synthesisers, where audio signals are kept in the analogue
domain, and control signals originate from digital representations
and are transformed into voltage control via a number of 
digital-to-analogue converters. This allows some level of 
interconnectivity via so-called modulation matrices that mimic
the modular approach, albeit in a smaller scale.

\section{Conclusions}

This paper attempted to demonstrate the usefulness of an analogue computing approach
to electronic and computer music research. It provided a general introduction
to the area, alongside tracing a parallel to modular voltage-controlled synthesisers.
Examples were given that had direct relevance to analogue audio signal
processing, demonstrating some immediate applications in research and music
production. 

A survey of the state-of-the-art in analogue computing provided us with some
first candidates as technologies that might be ready for use. In fact, in one case
some interesting results had already been presented. Challenges remain, however,
given that the target outputs of analogue computing for music applications have
some key constraints of quality, including low signal-to-noise ratios and pitch/voltage
stability. Assessing this should play an important part in any future research. 

Another aspect that was raised was to do with programmability. We see that 
key developments in the area are necessarily linked to the potential 
of music programming systems. Having analogue computing-dedicated
music tools, similar to what exists for standard digital computers, will
play an important part of making the technology available to a wider range of 
users (musicians, artists). This possibly points out to another fertile field of computer
music research.

\end{document}